\documentclass[draft]{agujournal2019}
\usepackage{url} 
\usepackage{lineno}
\usepackage[inline]{trackchanges} 
\usepackage{soul}
\usepackage{graphicx}
\usepackage{graphics}
\draftfalse

\journalname{a journal}

\begin{document}

%
%


\title{Machine Learning Solar Wind Driving Magnetospheric Convection in Tail Lobes}

%
%




\authors{Xin Cao\affil{1}\thanks{now in University of Colorado Boulder}, Jasper S. Halekas\affil{1}, Stein Haaland\affil{2}, Suranga Ruhunusiri\affil{1}, Karl-Heinz Glassmeier\affil{3}}

 \affiliation{1}{Department of Physics and Astronomy, University of Iowa, IA, USA.}
 \affiliation{2}{Birkeland Centre for Space Science, University of Bergen, Bergen, Norway.}
 \affiliation{3}{Institut für Geophysik und Extraterrestrische Physik, Technische Universität Braunschweig, Braunschweig, Germany.}





\correspondingauthor{Xin Cao}{xin-cao@uiowa.edu}




\begin{keypoints}
\item The ARTEMIS and Cluster missions observe plasma convection in the magnetotail lobes in the far-downstream and near-Earth tail regions.
\item Magnetospheric convection can be driven by the solar wind and the associated magnetospheric activity.
\item Machine learning models can be used to study the potential external driving mechanism of the magnetospheric convection.
\end{keypoints}

%
%

%
%


\begin{abstract}

To quantitatively study the driving mechanisms of magnetospheric convection in the magnetotail lobes on a global scale, we utilize data from the ARTEMIS spacecraft in the deep tail and the Cluster spacecraft in the near tail. Previous work demonstrated that, in the lobes near the Moon, we can estimate the convection by utilizing ARTEMIS measurements of lunar ions’ velocity. In this paper, we analyze these datasets with machine learning models to determine what upstream factors drive the lobe convection in different magnetotail regions and thereby understand the mechanisms that control the dynamics of the tail lobes. Our results show that the correlations between the predicted and test convection velocities for the machine learning models (>0.75) are much better than those of the multiple linear regression model (~0.23-0.43). The systematic analysis reveals that the IMF and magnetospheric activity play an important role in influencing plasma convection in the global magnetotail lobes.
\end{abstract}


%
%

%


%
%
%
%

\section{Introduction}
Characterizing the plasma convection in Earth’s tail regions is important to help us understand the global magnetospheric dynamics. \cite{Haaland2008, Haaland2009} utilized Cluster data \cite{escoubet_cluster_1997} to show that the plasma convection at ~10 RE downtail has opposite lateral patterns in the southern and northern lobes. For instance, the convection shows a pattern such that the north-south convection moves towards the current sheet in the magnetotail. \cite{Ohma2019} revealed that the asymmetry of the convection flow could also be affected by magnetic reconnection in the tail, which relates to magnetospheric activity. \cite{Cao2020b} used the two Acceleration, Reconnection, Turbulence, and Electrodynamics of Moon's Interaction with the Sun (ARTEMIS) lunar ion data \cite{angelopoulos_artemis_2011} to show that the dawn-dusk component of plasma convection velocity near the Moon’s orbit (~60 RE) has a high correlation with the corresponding component of the upstream Interplanetary Magnetic Field (IMF). The magnetosphere of the Earth responds to the solar wind flow and IMF, through the Dungey Cycle driven by dayside magnetic reconnection \cite{Dungey1961}. Based on previous studies, the magnetospheric plasma in both lobes tends to move towards the central plasma sheet. \cite{kissinger2011steady} showed that variations in solar wind conditions can also control the magnetospheric convection. The magnitude of the convection velocity is influenced by the upstream solar wind conditions, e.g. solar wind dynamic pressure, the IMF and its clock angle, and the magnetospheric activity, as measured using the Dst (disturbance storm time) index.

The Moon has a tenuous exosphere, which is mainly composed of neutrals sourced from the surface of the Moon via various processes, e.g. solar wind sputtering, and thermal and chemical release \cite{Stern1999, Sarantos2012, Cook2013, vorburger2014first}, and micrometeorite impact \cite{Hartle2006, Halekas2011,horanyi2015permanent}. Some of the neutral particles can be eventually transformed to same-mass heavy ions through photoionization, charge exchange, and electron impact ionization \cite{mcgrath1986sputtering, Sarantos2012, huebner2015photoionization, Zhou2013}. The motion of lunar ions in the magnetotail lobes reflects their interaction with the exosphere, the lunar surface, and the ambient environment of the tail lobes.

Compared to other regions such as the solar wind or the magnetosheath where the ambient plasma density is much higher than that of the lunar ions \cite{Halekas2011,chu2021electrostatic}, the lunar ion density in the terrestrial magnetotail lobes is comparable to or even larger than that of the ambient lobe plasma, and the background flow is commonly sub-Alfvénic \cite{Halekas2018,liuzzo2021investigating}. Therefore, the magnetospheric tail lobes are a unique environment in which to study the dynamics of the lunar ions. In this scenario, \cite{cao2019interaction,Cao2020a} analyzed ARTEMIS measurements to find that lunar ions are predominantly accelerated by magnetic tension and pressure forces. As a consequence of this process, the lunar ions are eventually coupled to the ambient plasma convection by the mass loading effect. Accordingly, the plasma convection in the deep magnetotail lobes can be estimated by measurement of lunar ion motion. This technique allows ARTEMIS to estimate the convection velocity by utilizing the heavy lunar ions (which have higher energy per charge for a given convection speed), despite the fact that it cannot typically directly detect the convection of ambient low-mass ions, given the large positive spacecraft potential in the tenuous lobe environment \cite{Cao2020b,cao2020influence}.

In contrast, the convection velocity in the near-Earth tail regions can be directly measured by the Electron Drift Instrument (EDI) on the Cluster spacecraft. The EDI emits electron beams, and detect their return to the spacecraft after one or more gyrations. By continuously tracking the emitted beam, the electron gyro center drift and thus the convection can be monitored \cite{paschmann1997electron}. The EDI measurement of convection velocity is not affected by the low density of ambient plasma \cite{Haaland2008}, and the measurement technique has been extensively validated in the magnetotail lobes \cite{Noda2003, Haaland2007, Haaland2008}.

Over the past few years, machine learning techniques have been widely used in space physics and planetary science \cite{camporeale2018machine}. For instance, \cite{kerner2019novelty} utilized a machine learning algorithm to detect novel geologic features in multispectral images of the Martian surface. \cite{wagstaff2019enabling} used machine learning methods to study thermal anomalies, compositional anomalies, and plumes of icy matter from Europa's subsurface ocean. \cite{nguyen2019automatic} utilized machine learning techniques to automatically detect the terrestrial bow shock and magnetopause from in-situ data, and \cite{lazzus2017forecasting} used machine learning algorithms to forecast the Dst index of the Earth. \cite{kronberg2020prediction} utilized machine learning techniques to study solar wind control of energetic particles and X-ray.

In this paper, we used lunar ion data from the ARTEMIS spacecraft to infer lobe plasma convection velocity near the Moon in the deep magnetotail lobes (\~~60 RE), EDI measurements from Cluster to determine the convection velocity in the near-Earth magnetotail between ~ 10 – 15 RE, and solar wind and Dst index data from NASA’s OMNI data set. We analyzed the relationship between magnetotail convection and upstream solar wind conditions and magnetospheric activity index. Based on the outputs of two ensemble learning methods: Random Forest and Gradient Boosting Decision Tree (GBDT), the results confirm that the lobe plasma convection in the tail regions is controlled by upstream solar wind and magnetospheric activity.

\section{Magnetotail Observations and Model Methods}

The two spacecraft of the ARTEMIS mission, P1 and P2, have been orbiting the Moon since mid-2011. We utilized measurements from two of the on-board instruments: Electrostatic Analyzer (ESA) \cite{McFadden2008} and Flux Gate Magnetometer (FGM) \cite{auster2008themis}. The ESA measures the ion distribution for energies between a few eV and 25 keV and the electron distribution for energies between a few eV and up to 30 keV \cite{McFadden2008}. The FGM measures the vector magnetic field at a cadence of ~4 sec minimum. The four Cluster spacecraft fly in formation in a high inclination 4x20 RE polar orbit, with apogee in the magnetotail between August - November. From Cluster,  we use EDI measurements of the convection in the tail lobes \cite{paschmann1997electron}. Since the measurement regions of the two missions extend from near Earth to the Moon’s orbit in the magnetotail, the coordinate system utilized in this study is Geocentric Solar Magnetospheric (GSM), in which the +XGSM axis is defined to be oriented towards the Sun from the center of the Earth, the +ZGSM axis towards the direction such that XZ plane contains the geomagnetic dipole axis, and the +YGSM axis completes the right-handed system. The data utilized in this study contains over 240,000 Cluster data points in the near-Earth tail regions and over a few thousand ARTEMIS data points in the deep tail lobe regions near the Moon’s orbit. The near-Earth measurement by the Cluster spacecraft was limited to be within -8 RE < YGSM < +8 RE and +8 RE < ZGSM < +16 RE for the northern lobe and be within -8 RE < YGSM < +8 RE and -16 RE < ZGSM < -8 RE for the southern lobe \cite{Haaland2008}. The measurements made in the deep tail lobes typically reveal a dominant positive or negative Bx component, which indicates the northern or southern lobes, respectively. As discussed in \cite{Cao2020b}, since the mass loading effect couples the lunar ions to the ambient plasma convection, the motion of lunar ions serves as an approximate tracer of the convection patterns in the tail lobes. More details about this method can be found in \cite{Cao2020a,Cao2020b,cao2020influence,cao2021using}.

The machine learning techniques we utilized in this study include random forest and GBDT, which are ensemble learning models \cite{friedman2001elements}. The basic principle of this category of model is to integrate a group of weak learners into a strong learner in order to obtain a better performance, where the weak learner represents a single decision tree in the models. In contrast to a weak learner that performs at least better than random guessing, a strong learner is a model doing the prediction work as well as possible compared to the test dataset. Both random forest and GBDT utilize a data structure called decision tree as the substructure of their models. The random forest and GBDT models usually construct a number of decision trees. Each tree in the random forest model receives a sub-dataset as the training dataset, which is sampled randomly with the replacement from the whole training dataset. This resampling method is termed the Bootstrap method in statistics \cite{friedman2001elements}. The key difference between the two models is that the GBDT model focuses more on the trees with larger errors during each iteration of the training process. Both the random forest and GBDT models have good ability to reduce the overfitting problem in machine learning and have been widely used in prediction and/or regression in scientific problems. The inputs of the models in this study include the solar wind IMF vector, the solar wind dynamic pressure, the local measured Bx, the Dst index, and the spacecraft locations in the magnetosphere. The Dst index describes the magnetospheric activity which can be used as an upstream monitor indirectly. The outputs of the models include the lateral components of convectional velocity: Vy and Vz. The combined training data are randomly extracted from 80\% of the dataset of each mission, and other 20\% of the dataset are used as the test dataset. During the training process, we also used 10-fold cross-validation \cite{friedman2001elements} to help further reduce the potential overfitting, and the optimal hyper-parameters in the models are obtained by using the grid search method. Because of the parallel computing ability, the random forest model was trained by using 8 threads simultaneously.

\section{Results of Random Forest and GBDT models}

The lateral component of magnetospheric convection in the lobes at different downtail distances has been shown to correlate with the upstream solar wind IMF direction \cite{Haaland2008,Haaland2009,Cao2020b}. The observed magnetospheric asymmetries such as dusk-dawn shift of the polar cap boundary and auroral zone flow speed asymmetry can affect the open field lines in the lobes, which may be controlled by the upstream dawn-dusk IMF \cite{Cowley1981}. \cite{Tenfjord2015, Tenfjord2018} utilized an MHD model to conclude that the driving mechanism might be due to the upstream lateral magnetic flux transferring to the nightside of the Earth’s magnetosphere, which then affects the plasma convection in the magnetotail lobes. Convection in the Z direction is strongly influenced by dayside solar wind-magnetosphere coupling \cite{dungey1963interactions}. This dayside coupling is also reflected in magnetospheric disturbance indices like e.g., the Dst index, so a correlation between lobe convection and the Dst index is often observed \cite{Haaland2009}, and a larger Dst implies lower activity and thus lower convection in the tail region.

\begin{figure}
\centering
\includegraphics[]{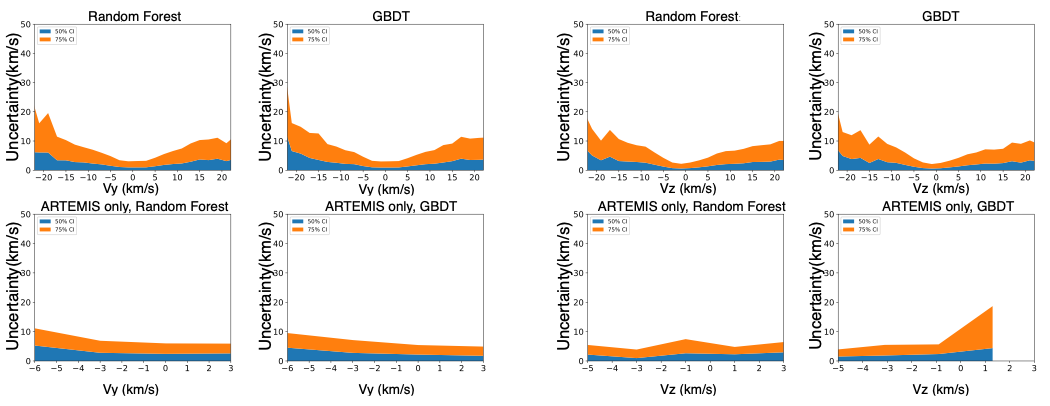}
\caption{The 50\% and 75\% confidence intervals of the uncertainties (see the text) for the random forest and GBDT models respectively for Vy and Vz components of magnetospheric convection and for the near-Earth and deep magnetotail lobes. The left four panels (a) show the Vy’s and the right four panels (b) show the Vz’s.}
\end{figure}

Figure 1 illustrates the uncertainties of Vy and Vz convection values predicted by the random forest and GBDT models, trained using the data set we described in the second section. The left four panels (1a) show the predicted Vy values of the random forest and GBDT models respectively for the combined data set from the two missions, and for the ARTEMIS results alone. The blue and orange colors represent the 50\% and 75\% confidence intervals (CI), with the uncertainties having corresponding probabilities less than the upper boundary of the shaded regions \cite{Ruhunusiri2018}. In general, the uncertainty of positive Vy is smaller than that of negative Vy. The uncertainty of Vy within 50\% CI is relatively small. However, the uncertainty of Vy within 75\% CI reveals somewhat larger asymmetry between positive and negative values, particularly when the velocity magnitude is larger than ~15 km/s. This is probably because the proportion of the velocities with a large magnitude is much smaller than that of the velocities with a smaller magnitude, which represent the majority of the convection velocity values in the magnetotail lobes \cite{Haaland2008}. The relatively small proportion of larger velocity values in the data set could result in a larger bias, increasing the uncertainty of the corresponding range. 

The predictions of Vy by the random forest model are better than those from the GBDT model. For the prediction of the models for the combined data set, the overall relative deviation of 50\% CI’s uncertainty is around 0.25, and that of 75\% CI’s uncertainty for the only ARTEMIS data could be up to 0.7-1.0. We will show that the predictions of these two models are much better than that of a traditional multiple linear regression model in the next section. The right four panels (1b) correspondingly show predicted Vz values for the random forest and GBDT models. As the figure shows, compared to the Vy prediction, the overall prediction of Vz by the GBDT model for the combined data set are slightly better than that of the random forest model. However, the GBDT model’s prediction for the ARTEMIS measurement appears a larger uncertainty for the positive Vz. Besides, the predictions of Vy and Vz from both models for the combined data set are better than those for the ARTEMIS data alone, likely due to a higher uncertainty in the measurement of magnetotail convection using the ARTEMIS lunar ion data.

\begin{figure}
\centering
\includegraphics[]{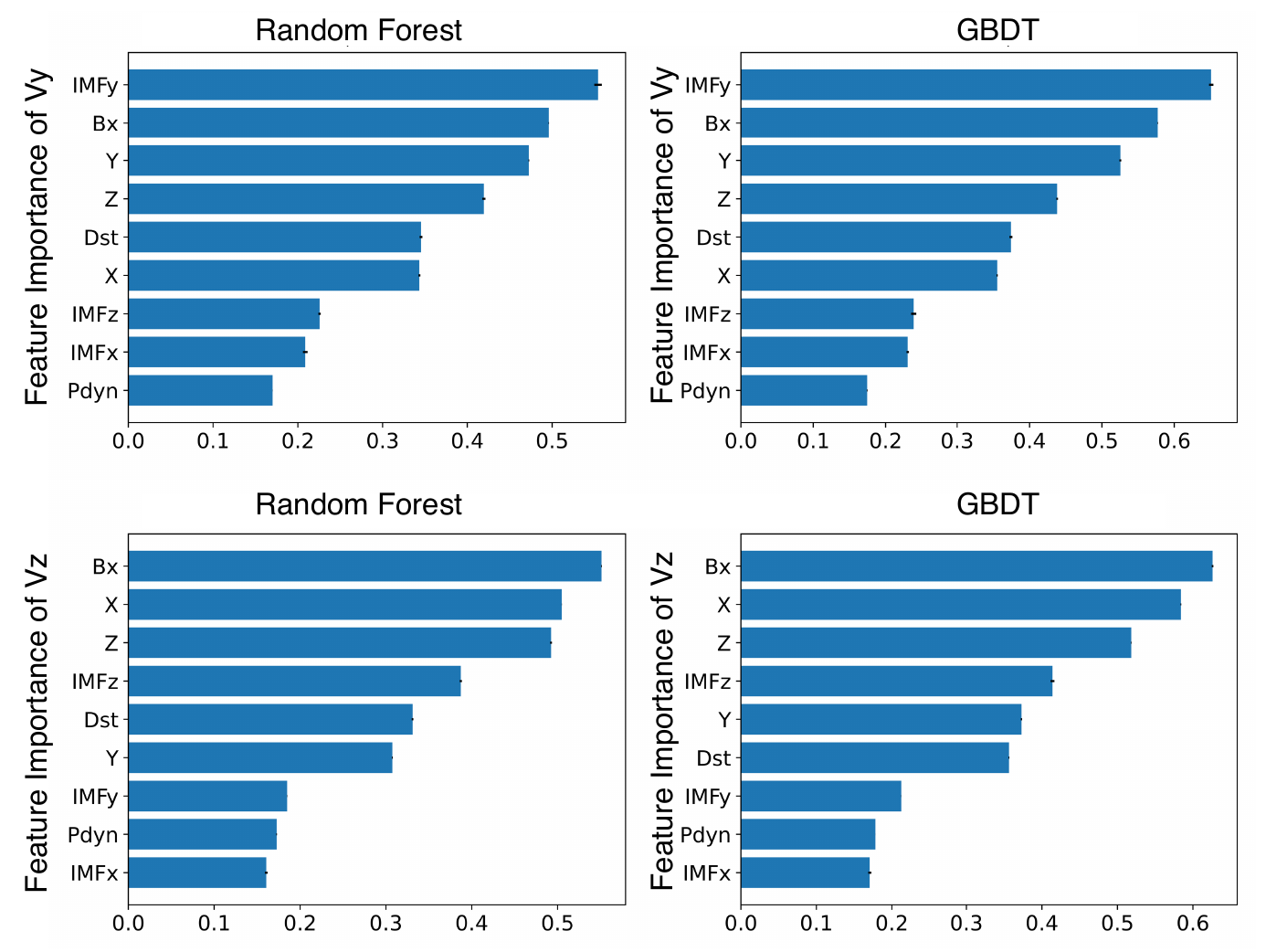}
\caption{The feature importance of the velocity prediction for the random forest and GBDT models. The upper panels are for the Vy component of magnetotail convection and the lower panels are for Vz, respectively for the random forest and GBDT models. The error bar in black represents the standard deviation of each parameter’s feature importance. The measurement locations of the spacecraft are represented by the three dimensional coordinates X, Y and Z. The IMFx, IMFy and IMFz mean the different components of the interplanetary magnetic field. The Dst index and Bx represent the magnetospheric activity index and the X component of local magnetic field in the tail lobes. The Pdyn represents the dynamic pressure of the upstream solar wind.}
\end{figure}

In order to identify which input parameters are more influential in controlling the plasma convection in the magnetotail, we show the feature importance of each input parameter for the convection velocity prediction, as depicted in Figure 2. In this study, the feature importance is defined to be the score decrease of the model when randomly shuffling the values of each single feature, which indicates how much the model is dependent on that feature. We calculated 10 random shuffles for each parameter and calculated their average, in order to reduce the potential bias resulting from a single random shuffle. The results of random forest and GBDT models both reveal that the upstream IMF By has the highest feature importance for the Vy component, which indicates that it plays the biggest role of the chosen input parameters in driving the lobe convection in the magnetotail, consistent with the previous observations \cite{Ohma2019} and \cite{Case2018, Case2020}. Next, the feature importances of the local magnetic field and the Dst index are also relatively significant, which is probably linked to the fact that physical processes in the near-Earth magnetosphere, as indicated by the geomagnetic activity, have a pronounced effect on physical processes in the downstream region, due to the global disturbance of the magnetic flux transport through the magnetospheric Dungey Cycle. In contrast, the geometric locations and local magnetic field in the magnetotail lobes hold the highest feature importance for the Vz component of the convection. The Vz dynamics may be more strongly affected by the different magnetic field structure between the near-Earth and far-tail regions. The local measured Bx also plays an important role in controlling the Vz component, since its sign differs between the two lobes, with the northward-southward convection generally towards the central current sheet \cite{Haaland2008,Cao2020b}. The feature importance of the Dst index and IMFz could also be comparably important since the geomagnetic activity and the IMFz can influence the convection velocity Vz’s pattern \cite{Haaland2008, Haaland2009, Ohma2019}.

\section{Correlation Analysis}

As described in the previous sections, the magnetospheric convection in the tail lobes is largely driven by the upstream solar wind conditions and magnetospheric activity. In this section, we split the two lobes and statistically investigate the correlation between solar wind conditions and magnetospheric activity, and the plasma convection of the lobes at different down-tail distances. The correlation calculation is made by restricting the data between the 10\% and 90\% percentiles, in order to reduce biases from outliers.

\begin{table}
\centering
\includegraphics[]{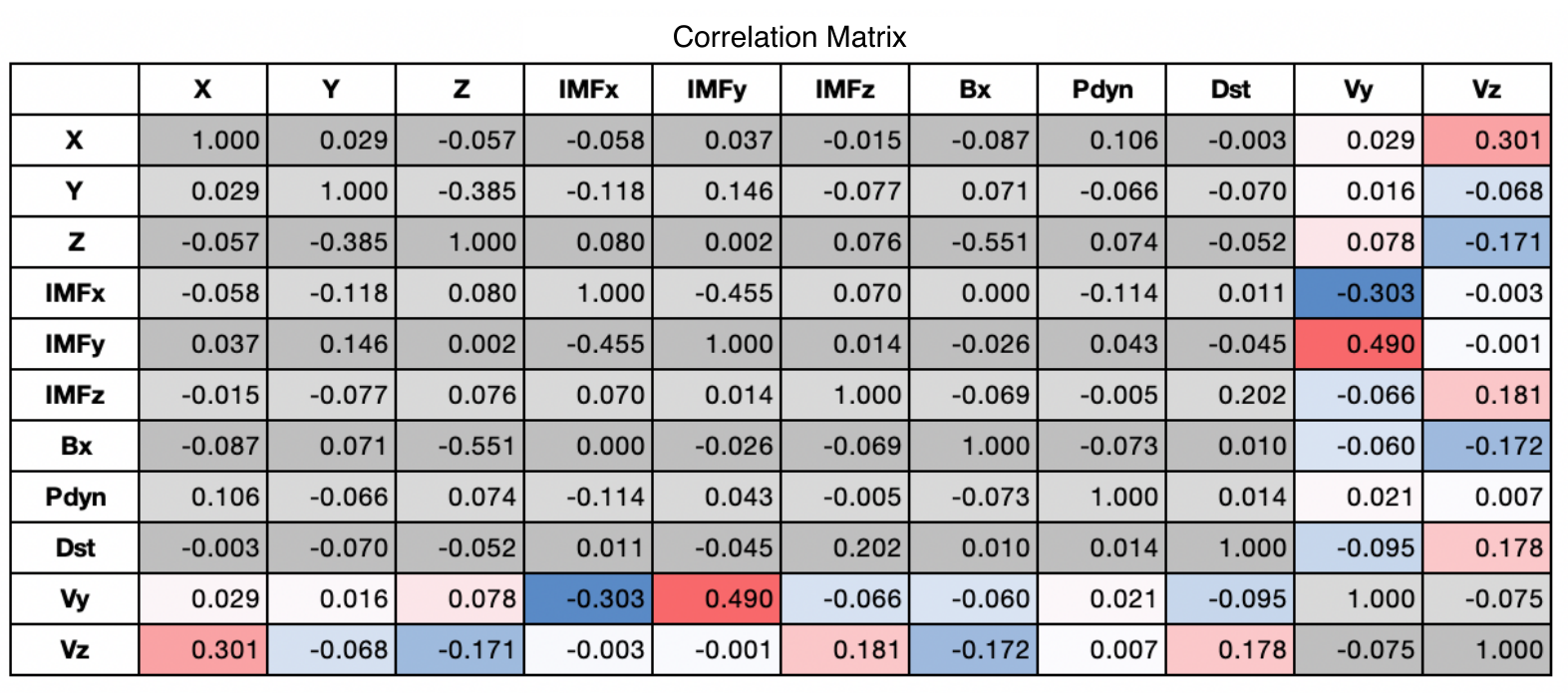}
\caption{The correlation matrix between the upstream conditions and the lateral components of the convection velocity in the northern lobes throughout different tail lobe regions. The colored cells represent the correlation coefficients between each of upstream conditions and the Vy and Vz of the magnetospheric convection. Red colors indicate larger correlations and blue colors indicate larger anticorrelations.}
\end{table}

Table 1 shows the correlation coefficients between each of the upstream parameters, the geometric locations of the measurement made in the magnetosphere, and the lateral velocity components of the magnetospheric convection in the northern lobe measured by Cluster and ARTEMIS. The correlation between the convection Vy and the IMF By has the largest value (~ 0.5) in the northern lobe. Corresponding, they have the largest anti-correlation in the southern lobe. This is consistent with previous observations that showed that the dawn-dusk convection in the tail lobes is strongly controlled by the IMF By \cite{Haaland2008, Case2018, Case2020, Ohma2019, Cao2020b}. The correlations between northern and southern lobes (not shown here) are similar but not exactly symmetric, which is probably due to two reasons: (1) the magnetic tilt of the Earth results in the structural asymmetry of the two lobes relative to the incident solar wind; (2) the measurement locations in the magnetosphere are not equally distributed among different lobes. The IMF By that affects the geometry of the upstream interaction with the solar wind can influence the asymmetry of the convection in the magnetosphere \cite{Tenfjord2015}.

Compared to the Vy component, the convectional Vz has a relatively significant correlation coefficient with the geometric location, which is consistent with our observation of the models. As discussed in previous section, this might be due to the influence of different magnetic field structures among different tail regions. The IMF Bz and the Dst index have comparable correlations with Vz in the two lobes, which is probably linked to the upstream reconnection-related interaction driving the magnetospheric convection, e.g. Dungey Cycle.  In addition, the lateral convection of the lobes is not significantly correlated with the upstream solar wind dynamic pressure, which is consistent with previous studies \cite{Haaland2008, Haaland2009}. In the global scale, the north-south component of plasma convection in the lobes (Vz) has different driving characteristics from that of Vy. For instance, the response of Vy and Vz to the Dst index appears different between the two tail lobes. The correlation of the geometric locations of the magnetosphere with Vz has also a relatively higher value compared to that with Vy, which might be because the northward-southward convection is affected by the variation of magnetic field structure from the near-Earth to the deep tail regions. 

Finally, we calculated the correction between the predicted and test velocities for the two machine learning models and the multiple linear regression model. The corresponding correlation for predicted and test Vy components of random forest is 0.76, and that of GBDT is 0.75, compared to a much smaller value of 0.23 from multiple linear regression. The corresponding correlation for predicted and test Vz components of random forest is 0.78, and that of GBDT is 0.77, compared to a value of 0.43 from multiple linear regression. The comparison of the correlation coefficients between these different models confirms that the machine learning models significantly outperform the multiple linear regression.

\section{Summary}

In conclusion, we investigated the potential driving mechanism of the plasma convection in the magnetospheric lobes with respect to upstream solar wind conditions and geomagnetic activity, by utilizing two types of machine learning models: random forest and GBDT. We used data from the ARTEMIS and Cluster missions, and from the OMNI dataset. This study indicated that the machine learning technique could be a useful tool to predict the response of the magnetospheric convection in the tail lobes to the upstream conditions, and revealed the feature importance of each potential driving parameter, with results that appear consistent with previous studies of the convection of the tail lobes \cite{Haaland2008, Haaland2009, Ohma2019, Cao2020b}. The ARTEMIS-Cluster-OMNI data-driven models demonstrate that the convection throughout the near-Earth and far-tail regions is largely controlled by the upstream solar wind parameters and as reflected also in magnetospheric activity indices. The IMF By values have significant correlations with the corresponding component of the convection in the tail lobes, as predicted. The geometric locations in the magnetosphere affect the Vz component more significantly than the Vy component. In addition, the Vz value in the tail lobes has a comparable response to the IMF Bz and the Dst index, which may indicate that the upstream solar wind driving mechanism consistently influences the geomagnetic environment in the near-Earth magnetosphere and the downstream tail regions. This is probably associated to the global plasma dynamics, e.g. Dungey Cycle. How the dynamics of other magnetospheric regions (e.g. plasma sheet or current sheet) respond to the upstream solar wind should be addressed in future studies, as it may help build a more complete picture of the solar wind – terrestrial magnetosphere coupling processes. The method of using machine learning techniques to study the magnetospheric convection could potentially be applied to the global magnetospheres of the Earth \cite{liu2012dipole} and other planets \cite{Cao2017} such as those of Mercury, Saturn, Jupiter, Uranus \cite{cao2013multifluid,cao2014seasonal,cao20153d,cao20163d,cao2017diurnal,cao2018diurnal,cao2021asymmetric} and Neptune and the non-global magnetic environment such as that of Mars \cite{chu2021dayside}.

\begin{verbatim}


\end{verbatim}


\acknowledgments
We acknowledge support from the Solar System Exploration Research Virtual Institute, Lunar Data Analysis Program grant 80NSSC20K0311, and NASA contract NAS5-02099. Karl-Heinz Glassmeier is financially supported by the German Ministerium für Wirtschaft und Energie and the Deutsches Zentrum für Luft- und Raumfahrt under contract 50 OC 1403. All ARTEMIS data are publicly available at NASA’s CDAWeb (https://cdaweb.sci.gsfc.nasa.gov) and the ARTEMIS site (http://artemis.ssl.berkeley.edu). We acknowledge James P McFadden for Electrostatic Analyzer data. We acknowledge the OMNI data which were obtained from the GSFC/SPDF OMNIWeb interface at https://omniweb.gsfc.nasa.gov.


%
%



\bibliography{agusample}

%
%
%
%
%

\end{document}